\documentclass[12pt]{iopart}

\usepackage{iopams}
\usepackage{graphicx}
\begin{document}

\title[Rotating thin-disk galaxies through the eyes of Newton]{Rotating 
thin-disk galaxies through the eyes of Newton}

\author{James Q. Feng and C. F. Gallo}

\address{Superconix Inc., 2440 Lisbon Avenue, Lake Elmo,
MN 55042, USA}
\ead{info@superconix.com}   
\begin{abstract}
By numerically solving the mass distribution in a rotating disk
based on Newton's laws of motion and gravitation,
we demonstrate that the observed flat rotation curves for 
most spiral galaxies correspond to 
exponentially decreasing mass density from galactic center 
for the most of the part except within the central core and 
near periphery edge.
Hence, we believe the galaxies described with our model
are consistent with that seen through the eyes of Newton.
Although Newton's laws and Kepler's laws seem to 
yield the same results when they are applied to 
the planets in the solar system,
they are shown to lead to quite different results
when describing the stellar dynamics in disk galaxies.
This is because that Keplerian dynamics may be equivalent to 
Newtonian dynamics for only special circumstances, but not generally for 
all the cases. 
Thus, the conclusions drawn from calculations based on 
Keplerian dynamics are often likely to be erroneous 
when used to describe rotating disk galaxies.
\end{abstract}


\maketitle

\section{Introduction}
\begin{figure}[h]
\includegraphics[scale=0.45]{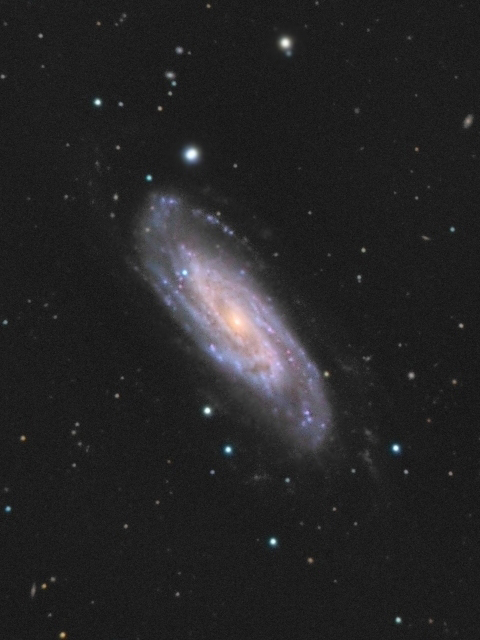}
\includegraphics[scale=0.45]{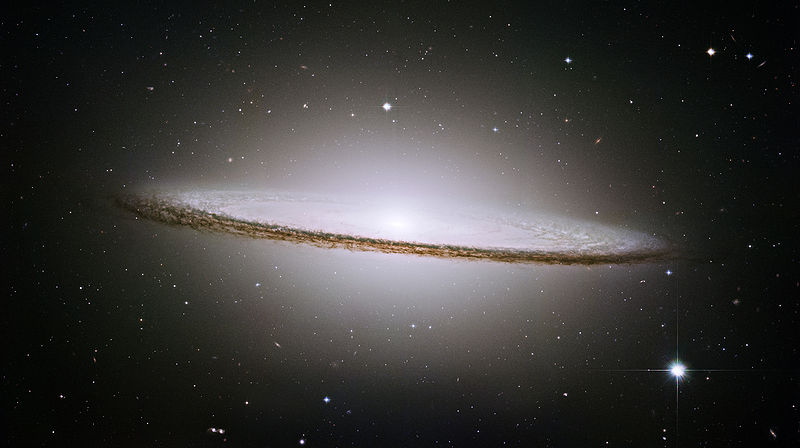}
\caption{\label{fig:galaxy}
Spiral galaxies: NGC 3198 and NGC 4594--also known as M 104 sombrero
galaxy.}
\end{figure}

A galaxy is a stellar system consisting of 
a massive gravitationally bound assembly 
of stars, an interstellar medium of gas and cosmic dust, etc.
Observations have shown that many (mature spiral) galaxies 
share a common structure with the {\em visible} matter
distributed in a flat thin disk (as in figure \ref{fig:galaxy}),
rotating about their center of mass in nearly circular orbits \cite{Binney87}.
Apparently, this typical behavior of galaxies 
is similar to that of our solar system with planets orbiting around the Sun
in a flat planetary plane.

For planets orbiting around the Sun,
Kepler's laws of planetary motion (obtained empirically) 
can provide accurate description.
Yet it was Isaac Newton, in his 
``Philosophiae Naturalis Principia Mathematica'',
who used mathematical expressions to 
show that Kepler's laws are consequences of 
Newton's laws of motion and universal law of gravitation \cite{Bennett07}. 
In addition, Newton found that Kepler's laws were only
part of the story of how objects move in response to gravity.
With his laws being discovered
by analyzing the orbits 
of planets around the Sun,
Kepler had no reason to believe that his laws would apply to 
other cases such as moons orbiting planets or comets orbiting the Sun.
But Newton was able to derive more general rules 
that can explain the motion of objects throughout the universe,
by analyzing his equations of gravity and motion.
Therefore, Newton's laws of motion and gravity have become
a crucial part of the foundation of modern astronomy,
whereas Kepler's laws can become misleading if not applied correctly
with sufficient care
to cases other than planets orbiting the Sun.
For example, Kepler's third law,
stating that more distant objects 
rotate around the center at slower average speeds,
cannot describe the typically observed {\em flat} orbital velocity
that remains invariant 
for most part of a galaxy outside its central core
\cite{Rubin70, Roberts75, Bosma78, Rubin80}. 

The fundamental difference between a galaxy and the solar system
is that the mass is apparently distributed across the entire galaxy 
whereas the solar system has its mass concentrated at 
the center in the Sun.
Each planet in the solar system can be quite reasonably treated as a
point mass moving in a spherically symmetric gravitational field
stemming from a large central point mass. 
The spherical symmetry of gravitational field in the solar system
greatly simplifies the mathematical analysis, 
because the gravitational potential at any position is 
basically determined by the distance from the center and the mass 
of the Sun, with contributions from other planets negligible. 
Actually, the treatment similar to that for the solar system
can be directly extended to the situation of 
distributed mass system as long as it retains the spherical symmetry,
except that now the equivalent mass at the center is not a constant but
equals to the mass enclosed 
by the concentric spherical surface through 
that point of interest.
According to Newton's first and second theorems,
if certain amount of mass is uniformly distributed in a spherical shell,
this shell exerts no net 
(gravitational) force on any mass at any point inside it but 
attracts any mass outside it as if its mass is concentrated 
at the center \cite{Binney87}.
Unfortunately, the thin-disk galaxies do not possess 
such a simplification-enabling spherical symmetry.
So, much more sophisticated mathematical treatments are needed
to {\em correctly} apply Newton's laws to the thin-disk galaxies. 

Here in this work, 
we demonstrate an effective numerical method for computing 
either mass density distribution for a given orbital velocity profile
or vice versa by solving
the governing equations based on Newton's laws for 
an axisymmetric thin-disk galaxy of finite size.
We also quantitatively illustrate the possible misleading results 
due to incorrect application of Kepler's laws to 
the same thin-disk galaxy.
In other words, we show that the observed behavior of disk galaxies 
can be described and explained by application of Newton's laws, 
but not by Kepler's laws that may only be regarded 
as a special case derived from Newton's laws
for spherically symmetric gravitational field.

\section{Governing equations}
\begin{figure*}
{\includegraphics[clip=true,scale=0.72,angle=90,angle=90,angle=90,viewport=50 50 450 800]{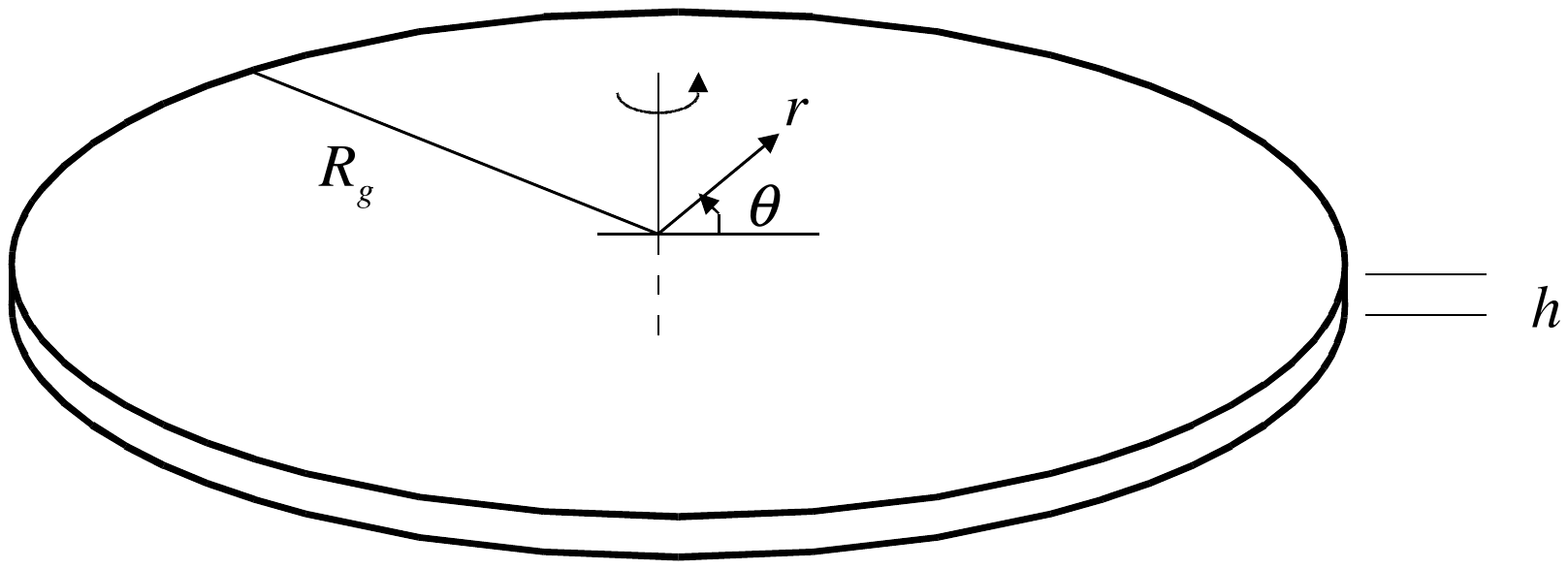}}
\caption{\label{fig:sketch}Definition sketch of the thin-disk model
considered in the present work.
The mass is assumed to distribute axisymmetrically in the circular disk
of uniform thickness $h$
with a variable density as a function of radial coordinate $r$
(but independent of the polar angle $\theta$).}
\end{figure*}

Because much of the mass of a galaxy resides in stars,
we can in principle compute gravitational field of 
a large collection of stars by adding the point-mass fields of all 
the stars together for any spot of interest. 
For convenience of mathematical treatment, however,
here we represent a galaxy by a continuum of 
axisymmetrically distributed mass in a circular disk of radius $R_g$
as shown in figure~\ref{fig:sketch}.
Thus, in the present model 
we consider continuous mass density distribution instead of 
discrete mass points scattered throughout the disk.
This kind of approximation is typically valid when the mass density in stars
is viewed on a scale that is small compare to the size of the galaxy,
but large compared to the mean distance between stars \cite{Binney87}.
Physically, the stars in a galaxy must rotate about the galactic center 
to maintain the disk shape.
Without the centrifugal effect due to rotation, the stars 
would collapse into the galactic center as a consequence of the 
gravitational field.
It is also reasonable to assume the galaxy is in an approximately
steady state with the gravitational force and centrifugal force 
balancing each other, 
in view of the fact that most disk stars have completed 
a large number of revolutions \cite{Binney87}.

Let's consider the force density on a test mass
at $(r, \theta = 0)$ generated by the gravitational attraction due to 
the summation (or integration) of 
a distribution of mass density $\rho(\hat{r})$ at position 
described by the variables of integration $(\hat{r}, \hat{\theta})$.
Here the distance between $(r, \theta)$ and 
$(\hat{r}, \hat{\theta})$ is
$(\hat{r}^2 + r^2 - 2 \hat{r} r \cos\hat{\theta})^{1/2}$
and the vector projection between the two points is
$(\hat{r} \cos\hat{\theta} - r)$.
Thus in a steady state, the mechanical balance between
the gravitational force 
(due to the summation of mass in a series of concentric rings)
and centrifugal force 
at every test point ($r$, $\theta = 0$) on the disk,
according to Newton's laws of motion and gravitation,
can be written as an integration equation 
(in terms of force per unit mass)
\begin{equation} \label{eq:force-balance0}
\int_0^1 \left[\int_0^{2 \pi}
\frac{(\hat{r} \cos \hat{\theta} - r) d\hat{\theta}}
{(\hat{r}^2 + r^2 - 2 \hat{r} r \cos \hat{\theta})^{3/2}}\right]
\rho(\hat{r}) h \hat{r} d\hat{r}
+ A \frac{V(r)^2}{r}
 = 0 \, ,
\end{equation}
where all the variables are made dimensionless
by measuring lengths (e.g., $r$, $\hat{r}$, $h$)
in units of the outermost galactic radius $R_g$,
disk mass density ($\rho$) in units of
$M_g / R_g^3$ with $M_g$ denoting the total galactic mass,
and velocities [$V(r)$] in units of the
characteristic galactic rotational velocity $V_0$
(as usually defined according to the problem of interest). 
The disk thickness $h$ is assumed to be constant and small
in comparison with the galactic radius $R_g$.
The results are expected to be insensitive 
to the exact value of this ratio
as long as it is small.
There is no difference in terms of physical meaning between
the notations $(r, \theta)$ and $(\hat{r}, \hat{\theta})$; 
but mathematically the former denotes the independent variables 
in the integral equation (for $\theta = 0$)
whereas the latter the variables of integration. 
The gravitational force represented as the summation of 
a series of concentric rings is described
by the first term (double integral) while the centrifugal forces
are described by the second term in (\ref{eq:force-balance0}).

Our process of nondimensionalization of the force-balance equation
yields a dimensionless parameter,
which we call the ``galactic rotation number'' $A$, as given by
\begin{equation} \label{eq:parameter-A}
A \equiv \frac{V_0^2 \, R_g}{M_g \, G} \, ,
\end{equation}
where $G$ ($= 6.67 \times 10^{-11}$ [m$^3$/(kg s$^2$)])
denotes the gravitational constant,
$R_g$ is the outermost galactic radius,
and $V_0$ is the characteristic velocity
(which may be equated here to 
the maximum asymptotic rotational velocity of a disk galaxy).
This galactic rotation number $A$
simply displays the ratio of
centrifugal forces to gravitational forces.
For typical galactic values of $R_g$, $V_0$, and $M_g$
we obtain $A \sim 1.6$ as will be shown in detail later.

When solving for the mass density $\rho(r)$ with $V(r)$ given,
we need to impose an overall constraint such that 
the total mass of the galaxy $M_g$ is
constant, namely,
\begin{equation} \label{eq:mass-conservation}
2 \pi \int_0^1 \rho(\hat{r}) h \hat{r} d\hat{r} = 1.
\end{equation}
This constraint due to the conservation of mass can actually
be used to determine the value of galatic rotation number $A$.

It is known that
the integral with respect to $\hat{\theta}$ in (\ref{eq:force-balance0})
can be written as \cite{Binney87} (pp. 72-73)
\begin{equation} 
\int_0^{2 \pi}
\frac{(\hat{r} \cos \hat{\theta} - r) d\hat{\theta}}
{(\hat{r}^2 + r^2 - 2 \hat{r} r \cos \hat{\theta})^{3/2}} 
= 2 \left[\frac{E(m)}{r (\hat{r} - r)} - \frac{K(m)}{r (\hat{r} + r)}\right]
 \, ,
\end{equation}
where $K(m)$ and $E(m)$ denote the complete elliptic integrals
of the first kind and second kind,
with
\begin{equation} 
m \equiv \frac{4 \hat{r} r}{(\hat{r} + r)^2} \, . 
\nonumber
\end{equation}
Thus, (\ref{eq:force-balance0}) becomes
\begin{equation} \label{eq:force-balance}
\int_0^1 \left[
\frac{E(m)}{\hat{r} - r} - \frac{K(m)}{\hat{r} + r}
\right]
\rho(\hat{r}) h \hat{r} d\hat{r}
+ \frac12 A V(r)^2
 = 0 \, .
\end{equation}
Equations (\ref{eq:force-balance}) and (\ref{eq:mass-conservation})
can be used to determine the mass density distribution 
$\rho(r)$ in the disk,
the galactic rotation number $A$, 
and subsequently the total galactic mass $M_g$,
all from measured values of $V(r)$, $R_c$, $R_g$ and $V_0$.
Seemingly complicated as it might be, these equations for $\rho(r)$
actually constitute a linear mathematical problem that guarantees 
uniqueness of solutions.
Conversely, these equations can also be used to determine 
the orbital velocity $V(r)$ if the mass density distribution $\rho(r)$
is known.
This is a well-defined mathematical problem completely deducible from the
available input data.

Because our governing equations are derived according to 
Newton's laws, they must be applicable to the solar system. 
As an example for $\rho(r) = \delta(r)/(\pi \, r)$ (namely, the
Dirac delta function in two-dimensional polar coordinates),
(\ref{eq:force-balance0}) or (\ref{eq:force-balance}) 
would yield 
\begin{equation} 
-\frac{1}{r} + A \, V(r)^2 = 0 \, ,
\nonumber
\end{equation}
which is exactly the familiar formula for so-called Keplerian velocity based on
Kepler's third law of planetary motion 
\begin{equation} \label{eq:Keplerian}
V_0 \, V(r) =  \sqrt{\frac{G \, M_g}{R_g \, r}} \, ,
\end{equation}
where $V_0 \, V(r)$ is the dimensional orbital velocity,
$M_g$ is basically the mass of the Sun, and $R_g \, r$ the 
dimensional distance from the Sun.

For a galaxy with mass distribution that is not spherically symmetric,
simple closed-form analytical solutions may not be tractable.  
Yet accurate numerical solutions can be computed with 
appropriately implemented computational techniques
as detailed in Appendix A. 
What it amounts to is nothing more than solving a 
linear algebra matrix problem using a well-established matrix solver.

\section{Mass distribution determined from given rotation curve}
\begin{figure*}
{\includegraphics[clip=true,scale=0.90,viewport=66 330 760 760]{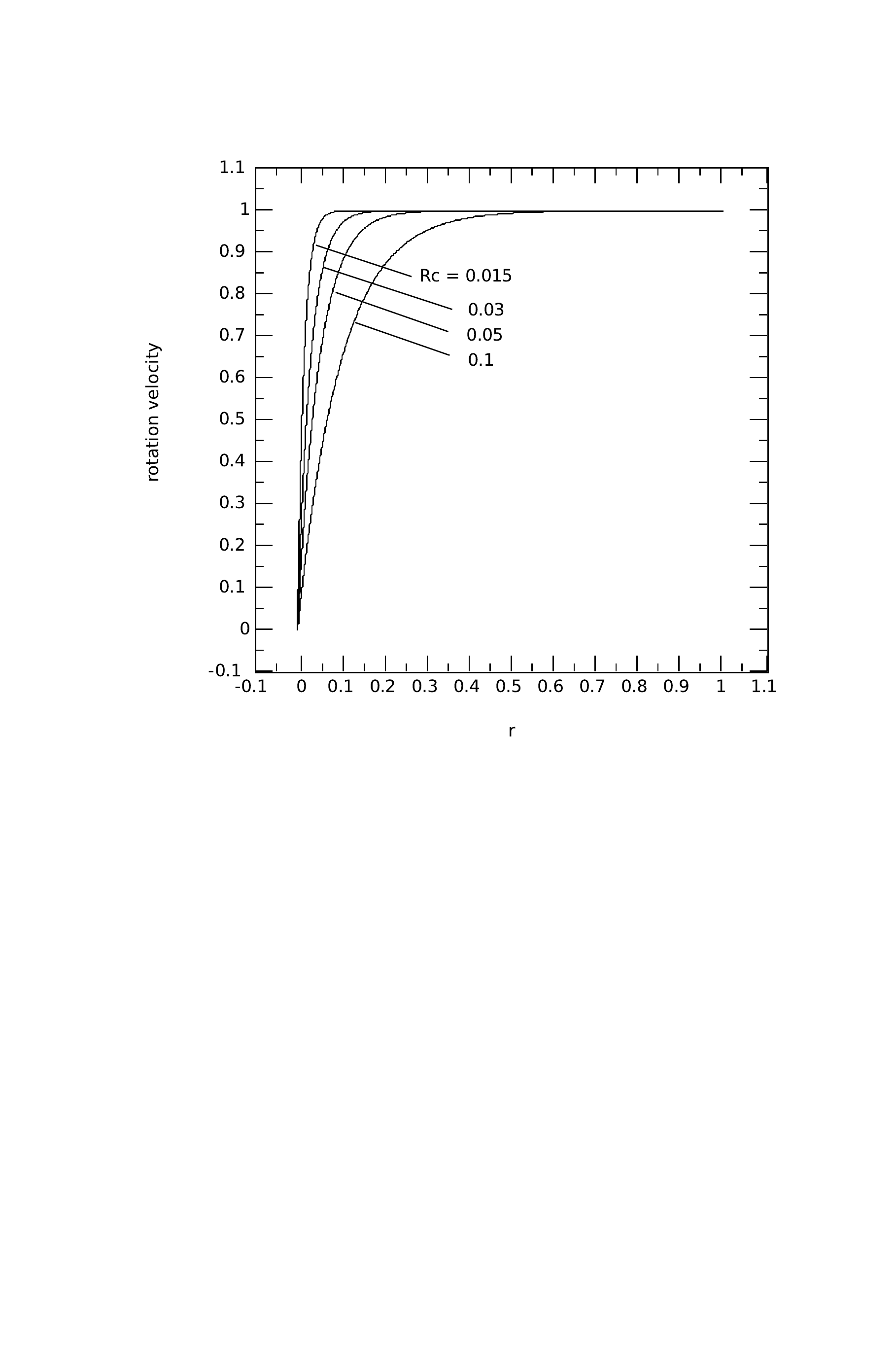}}
\caption{\label{fig:rotationcurve}Nondimensionalized 
orbital velocity profiles $V(r)$
according to the mathematically idealized formula
(\ref{eq:rotationCurves})
for $R_c = 0.015$, $0.03$, $0.05$ and $0.1$.}
\end{figure*}

The measurements of galactic rotational velocity profiles (also known as
``rotation curves'') of
mature spiral galaxies
reveal that
the rotation velocity typically rises 
linearly from the galactic center
(as if the local mass were in rigid body rotation) in a relatively small core,
and then with the slope decreasing continuously in a narrow transition zone it
reaches an approximately constant (flat) velocity extending
to the galactic periphery
\cite{Rubin70, Roberts75, Bosma78, Rubin80}.
These essential features may be mathematically idealized as\footnote{Because
the equations are solved numerically, the form of $V(r)$ can be almost 
arbitrary.  The idealized form of $V(r)$ presented here is just for 
convenience of illustrations rather than the limitation of 
mathematical solution techniques.}
\begin{equation} \label{eq:rotationCurves}
V(r) = 1 - e^{-r / R_c} \, ,
\end{equation}
where $V(r)$ denotes the (dimensionless) orbital velocity
(as measured in units of the maximum velocity in the flat part
that may be regarded as the characteristic velocity of 
galactic rotation $V_0$),
and $r$ the radial coordinate
from the galactic center
(in units of the outermost galactic disk radius $R_g$).
The parameter $R_c$ can be used to describe various radii of
the "cores" of different galaxies.
Close to the galactic center, namely when $r/R_c$ is small,
we have $V(r) \sim r/R_c$, describing a linearly rising rotation velocity, 
by virtue of the Taylor expansion
of $e^{-r/R_c}$.
Figure~\ref{fig:rotationcurve} shows
typical galactic rotation curves described by (\ref{eq:rotationCurves}).

With $V(r)$ given by (\ref{eq:rotationCurves}),
the mass density distribution $\rho(r)$ and the value of $A$
can be determined by computing solution to 
(\ref{eq:force-balance-residual}) and (\ref{eq:mass-conservation-residual}).
To compute numerical solutions,
the value of disk thickness $h$ must be provided;
we assume $h = 0.01$ (based on the measurements for the Milky Way galaxy).
By using $N = 1001$ nonuniformly distributed nodes
we found the obtained curves of $\rho(r)$ become reasonably smooth
and the values of galactic rotation number $A$ 
are discretization-independent.

\begin{figure*}[t]
{\includegraphics[clip=true,scale=0.90,viewport=66 330 760 760]{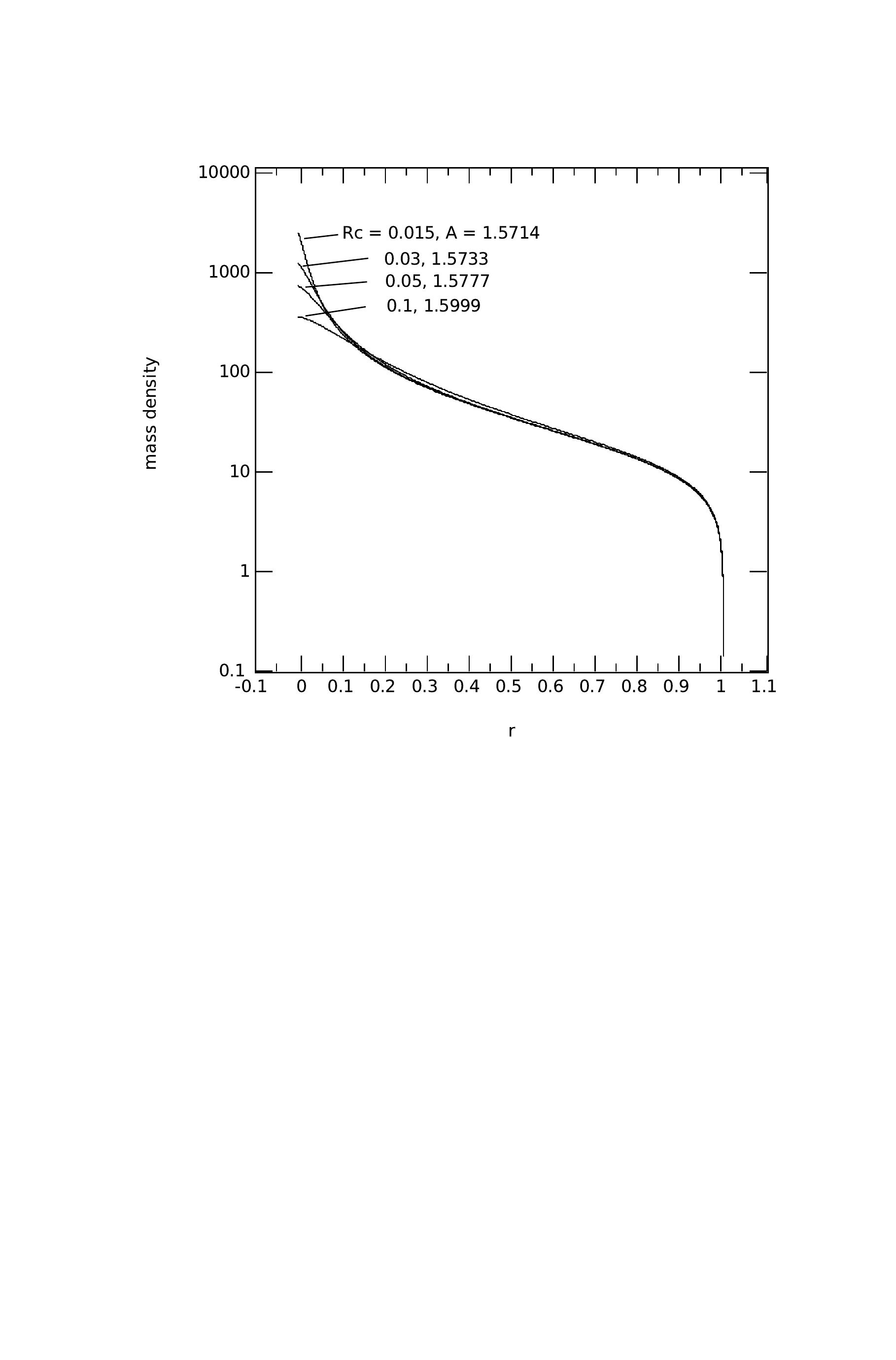}}
\caption{\label{fig:rho-r}The distributions of
mass density $\rho(r)$ computed based on Newtonian dynamics 
and given rotation curves for
$R_c = 0.015$, $0.03$, $0.05$, and $0.1$,
with $A = 1.5714$, $1.5733$, $1.5777$, and $1.5999$
determined as part of the numerical solutions.}
\end{figure*}

Shown in figure~\ref{fig:rho-r} 
are the computed mass density distributions $\rho(r)$ 
that satisfy the galactic rotation curves in figure~\ref{fig:rotationcurve}.
It is at the galactic center where mass attains the highest density.
Away from the galactic center, the mass density decreases rapidly
(with a slope becoming steeper for a tigher galactic core
indicated by smaller $R_c$).
However, beyond $R_c$, the mass density decreases rather 
gradually towards the galactic periphery
until reaching the galactic edge where it takes a sharp drop.
Noteworthy here is that the computed values of
galactic rotation number $A$ are within a small range around $1.6$
despite an order-of-magnitude variation of the galactic core radius $R_c$.

As apparent
in figure~\ref{fig:rho-r},
the computed $\log \rho$ decreases almost linearly with $r$
except for $r < 0.1$ (within the central galactic core)
and $r > 0.9$ (near the galactic edge).
This general feature is quite similar to the measured brightness
distributions (in typical spiral galaxies)
that are commonly fitted in an exponential form with regions of
central core and outer edge truncated.
For the case of $R_c = 0.015$ and $A = 1.5714$,
a least-square fit of our computed $\log \rho$ versus $r$
for $0.1 \le r \le 0.9$ yields
\begin{equation} \label{eq:log_rho}
\log \rho
= 5.4179 - 3.6802 \, r \, .
\end{equation}
This actually corresponds to an exponential function
$\rho = \rho_0 \, e^{-r/R_d}$ with
\begin{equation} \label{eq:rho_0Rd}
\rho_0 = 225.4 \mbox{ and }
R_d = 0.2717 \, ,
\end{equation}
which is known not to be able to generate the observed
flat rotation curves \cite{Freeman70, Binney87}.
Thus the much more rapid decrease of $\log \rho$ in the small
intervals [0, 0.1) and (0.9, 1]
must play important roles in compensating the deficiencies of the
simple exponential mass density distribution for matching the
commonly observed (flat) rotation curves.
If we assume the mass density follows roughly the exponential distribution
$\rho = \rho_0 \, e^{-r/R_d}$,
we would have the cumulative mass from the galactic center given by
\begin{equation} \label{eq:cumulative-mass}
2 \pi h \int_0^r \rho_0 e^{-\hat{r}/R_d}\hat{r}d\hat{r} = 
2 \pi h \rho_0 [R_d^2 - R_d (r + R_d) e^{-r/R_d}] \, .
\end{equation}
At $r = 1$, this yields a value of $0.922$ for $\rho_0 = 225.4$ and 
$R_d = 0.2717$,
indicating the exponential mass density distribution is likely to describe
about $90\%$ of the mass in a disk galaxy.
Another $\sim 10\%$ of the galactic mass seems to reside 
in the galactic core which may not follow the 
same exponential distribution (cf. figure~\ref{fig:rho-r}).

From the knowledge of $V_0$ and $R_g$ from measured
rotation curves, we can determine the value of $M_g$
based on computed value of the galactic rotation number $A$
(cf. (\ref{eq:parameter-A})) as
\begin{equation} \label{eq:M_g}
M_g
= \frac{V_0^2 \, R_g}{A \, G} \, .
\end{equation}
Because our computed $A$ varies little from $1.6$ for various $R_c$,
(\ref{eq:M_g}) suggests a general relationship of $M_g \propto V_0^2 R_g$ 
as what
Bosma \cite{Bosma78} 
found from evaluating mass versus size 
in a large number of observed disk galaxies.
In view of the fact that the values of $V_0$ are typically around 
$200$ (km/s), we then believe that the disk size of galaxies $R_g$
must be finite in order to keep the total mass of a galaxy $M_g$ from
becoming infinity. 
Suggesting finite disk size of galaxies does not 
necessarily mean that the mass density 
becomes zero outside the disk edge. 
We believe that 
the mass density beyond the galactic edge approaches the 
inter-galactic level of value and is roughly spherically symmetric,
which leads to inconsequential gravitational effect on 
rotation dynamics in the disk part.  

If we take the rotation curve
with $R_c = 0.015$
in figure~\ref{fig:rotationcurve} as
that of the Milky Way,
we have the galactic rotation number $A = 1.57$\footnote{Here we 
use the case of $R_c = 0.015$ only for the purpose of convenient illustration.
The value of $A = 1.57$ (with a variation of $1\%$)
is actually valid for a wide range of 
$R_c$ values as shown in figure~\ref{fig:rho-r}.}.
Then, from measured Milky Way values $V_0 = 2.2 \times 10^5$ (m/s)
and $R_g = 5 \times 10^4$ (light-years) $= 4.73 \times 10^{20}$ (m),
(\ref{eq:M_g}) yields
\begin{equation} \label{eq:M_g-value}
M_g
= 2.19 \times 10^{41} (\mbox{kg}) = 1.10 \times 10^{11} (\mbox{solar-mass}) \, .
\end{equation}
This value is in very good agreement with the Milky Way star counts
of 100 billion
\cite{Sparke07},
further including additional dust, grains, lumps,
gases and plasma in all galaxies.

With the given values of $M_g$ and $R_g$,
we can estimate the computed `radial scale'
$R_d \, R_g = 4.0$ (kpc) $= 1.234 \times 10^{20}$ (m) 
and the exponential disk central (surface)
mass density $\rho_0 h M_g/R_g^2 = 1340$ (solar-mass / pc$^2$)
based on (\ref{eq:log_rho}) for the Milky Way galaxy.
(Here, $1$ (pc) $= 3.086 \times 10^{16}$ (m) and 
$1$ (solar mass) $= 1.99 \times 10^{30}$ (kg).)
Compared with the results from fitting the brightness measurement data,
e.g., the radial scale of $2.5$ (kpc) and exponential disk central
brightness of $867$ (solar-luminosity / pc$^2$) \cite{Freudenreich98},
our computational results indicate 
either generally dimmer stars (than the Sun) or 
considerable amounts of cold gas exist
throughout the Milky Way with the total
mass density decreasing at slower rate than that of the
brightness (i.e., luminosity density).

Another example is the galaxy NGC 3198,
which has a (nearly idealized, often cited)
rotation curve with $V_0 = 1.5 \times 10^5$ (m/s),
$R_g = 30$ (kpc) $= 9.24 \times 10^{20}$ (m),
and $R_c \sim 0.015$.
Again with $A = 1.57$, we obtain $M_g = 1.98 \times 10^{41}$ (kg)
$= 9.9 \times 10^{10}$ (solar-mass).
As with the Milky Way, we can also predict the radial scale and
exponential central mass density for NGC 3198 as
$8.16$ (kpc) and $250$ (solar-mass / pc$^2$),
respectively (based on
$\rho_0 = 225.4$ and $R_d = 0.2717$ from (\ref{eq:log_rho}).
Compared with the radial luminosity profile
(which suggested an exponential disk with a radial scale of $2.63$ (kpc)
and central brightness $212$ (solar-luminosity / pc$^2$)
based on a total luminosity of $9.0 \times 10^9$ (solar-luminosity)
\cite{Begeman89},
our predicted mass density appears to decrease much more slowly
with stars (or mass objects) generally dimmer than the Sun.

Hence, the results computed here, as if they were obtained by Newton 
applying his laws of motion and gravitation
to solve the governing equations (\ref{eq:force-balance0})
and (\ref{eq:mass-conservation}), 
seem to be reasonably consistent with the observational measurements.  
In other words, the rotating thin-disk galaxies 
through the eyes of Newton are nothing more than 
massive gravitationally bound assemblies of objects  
governed by his same laws for the planet's motion in the solar system
albeit more sophiscated mathematical treatments are needed
to obtain the correct description 
than those with the planets in the solar system.

\section{Results based on Keplerian dynamics}
In the literature, 
many authors \cite{Bennett07, Sparke07, Volders59, Keel07, Rubin06, Rubin07} 
tend to simply apply 
Keplerian dynamics (which was derived from gravitational field 
generated by a spherically symmetric 
distribution of mass)
when analyzing the rotation behavior of thin-disk galaxies.
For example, Volders \cite{Volders59} demonstrated 
that spiral galaxy M33 does not spin as expected 
according to Keplerian dynamics--a result which was 
later extended to many 
other spiral galaxies \cite{Rubin70, Roberts75, Bosma78, Rubin80}.

Strictly speaking, a Keplerian potential 
(due to a point mass $M$ as that for the solar system)
is expressed as 
\begin{equation} \label{eq:Keplerian-potential}
\Phi(r)
= -\frac{G \, M}{r} \, ,
\end{equation}
For a distributed mass with spherical symmetry, 
the generalized form of Keplerian potential becomes
\begin{equation} \label{eq:Keplerian-like-potential}
\Phi(r)
= -\frac{G \, M(r)}{r} \, ,
\end{equation}
where $M(r)$ denotes the amount of mass enclosed by 
the concentric spherical surface of radius $r$ \cite{Binney87}.
Although (\ref{eq:Keplerian-like-potential}) comes from
the assumption of spherical symmetry,
it has often been used to determine the mass distribution and 
the total mass of a (disk) galaxy from a measured rotation curve, 
with $M(r)$ denoting the mass interior to 
radius $r$ from the galactic center.
For example,
in the recent versions of textbooks by Bennett {\it et al.} \cite{Bennett07}, 
Sparke and Gallagher \cite{Sparke07}, and Keel \cite{Keel07},
the value of $M(r)$ (also denoted as $M_r$ or $M(< r)$)
in a disk galaxy is 
simply determined from a known rotation curve $V(r)$ by
\begin{equation} \label{eq:mass-from-Keplerian}
M(r)
= \frac{r \, V(r)^2}{G} \, ,
\end{equation}
which has basically the same mathematical form as (\ref{eq:Keplerian}).

According to (\ref{eq:mass-from-Keplerian}), 
we would have an equation based on Keplerian dynamics 
for force balance as
\begin{equation} \label{eq:force-balance?}
2 \pi \, h \int_0^r \rho(\hat{r})\hat{r} d\hat{r}
- A \,r \,V(r)^2
 = 0 \, ,
\end{equation}
instead of (\ref{eq:force-balance0}). 
Here, the difference in mathematical forms between (\ref{eq:force-balance?})
and (\ref{eq:force-balance0}) should be quite clear. 
But whether there are much of practical differences between the two
may not be obvious. 
For example, authors like Sparke and Gallagher \cite{Sparke07}
and Keel \cite{Keel07} attempted to justify their usages 
of Keplerian formula for rotating thin disk galaxies by
stating that the result due to Keplerian formula does not differ 
more than $20\%$ from the actual result,
without showing a quantitative comparison. 
Therefore, we would like to 
examine a few comparative examples to see whether 
Keplerian dynamics (\ref{eq:force-balance?}) can be practically used as 
a close approximation to Newtonian dynamics
(\ref{eq:force-balance0}) for disk galaxies. 

For the orbital velocity $V(r)$ given by (\ref{eq:rotationCurves}), 
an analytical solution to (\ref{eq:force-balance?}) 
for $\rho(r)$ can be obtained 
as 
\begin{equation} \label{eq:solution-to-force-balance?}
\rho(r) = \frac{A}{2\pi\,h} \left[\frac{1}{r}
\left(1 - 2 e^{-r/R_c} + e^{-2 r/R_c}\right)
+ \frac{2}{R_c}\left(e^{-r/R_c} - e^{-2 r/R_c}\right)\right]
\, .
\end{equation}
It is not difficult to prove that $\rho(r) \to 0$ as $r \to 0$,
as in sharp contrast to that obtained according to Newton's laws
shown in figure~\ref{fig:rho-r}.
For $r >> R_c$ (i.e., outside the galactic core), 
(\ref{eq:solution-to-force-balance?} describes $\rho(r) \propto 1/r$,
as expected for the 
part where $V(r)$ is flat (constant) such that the 
first term in (\ref{eq:force-balance?}) becomes proportional to $r$.
In fact, many astronomers often consider 
the flat rotation curves, i.e., $V(r) = constant$, to indicate 
that the mass of a disk galaxy should continue 
to increase linearly well beyond its bright central region 
because $r\,V(r)^2 \propto r$ \cite{Rubin06, Rubin07}.

To satisfy (\ref{eq:mass-conservation}), the value of $A$ in 
(\ref{eq:solution-to-force-balance?}) is given by 
\begin{equation} \label{eq:solution-of-A}
A =
\frac{1}{1 - 2 e^{-1/R_c} + e^{-2/R_c}}
\, .
\end{equation}
When $R_c$ is small, e.g., $R_c = 0.015$, 
we have $e^{-1/R_c} \to 0$, e.g., $e^{-1/0.015} \sim 10^{-29}$;
therefore, $A \approx 1$ according to (\ref{eq:solution-of-A}).
This is in sharp contrast to the result based on formulas 
for disk galaxies (which predicts $A = 1.57$).  
If $A = 1$ were used in (\ref{eq:M_g}), we would have 
the total mass in Milky Way equal to 
$2.2 \times 10^{11}$ (solar-mass)--much more than that 
given by (\ref{eq:M_g-value}) and the Milky Way star counts.

\begin{figure*}[t]
{\includegraphics[clip=true,scale=0.80,viewport=66 330 760 760]{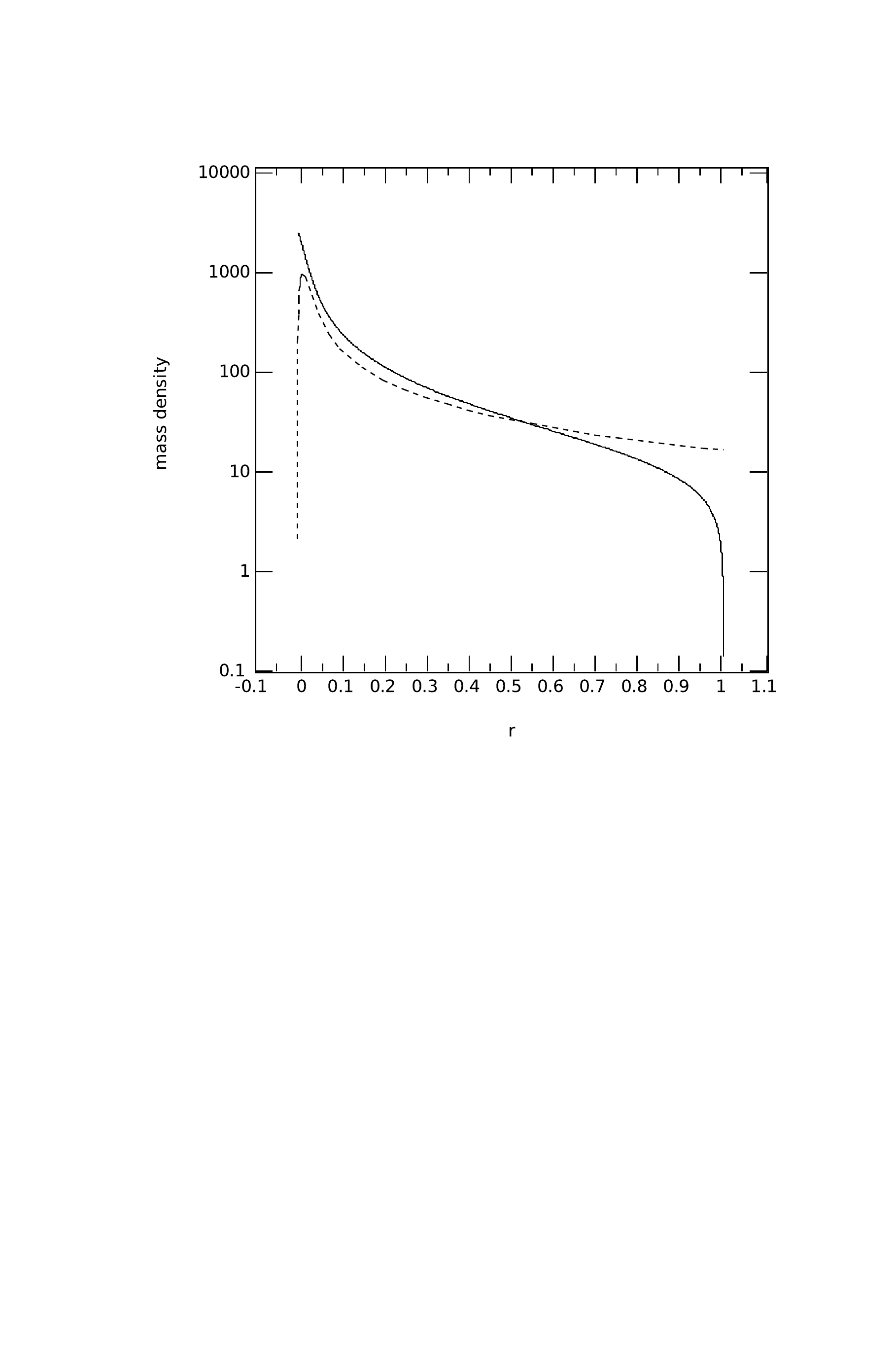}}
\caption{\label{fig:rho-comparison}The distribution of
mass density $\rho(r)$ computed based on Newtonian dynamics for
$R_c = 0.015$ (solid) and that derived based on Keplerian dynamics given by
(\ref{eq:solution-to-force-balance?}) with $A = 1$ 
and $R_c = 0.015$ (dotted).}
\end{figure*}

As a comparison, the distribution of $\rho(r)$ shown in 
figure~\ref{fig:rho-r} for $R_c = 0.015$ and that given by 
(\ref{eq:solution-to-force-balance?}) with $A = 1$ 
are shown in figure~\ref{fig:rho-comparison}.
Now, the differences between the two are obvious:
the Keplerian mass density behaves totally differently 
in the galactic core with $\rho(r) \to 0$ whereas that 
numerically obtained in \S~3 
has $\rho(r)$ monotonically decreasing with $r$ from 
its maximum value at galactic center to 
periphery.
Outside the galactic core,
the Keplerian mass density shows much slower decay
($\propto 1/r$) toward the galactic periphery
whereas that obtained numerically in \S~3
seems to be approximately exponential. 
Therefore, the mass density distribution determined 
based on Keplerian dynamics (\ref{eq:force-balance?}) 
cannot be a good approximation to
the actual $\rho(r)$ obtained numerically in \S~3
with Newton's laws strictly applied. 

\begin{figure*}[t]
{\includegraphics[clip=true,scale=0.80,viewport=66 330 760 760]{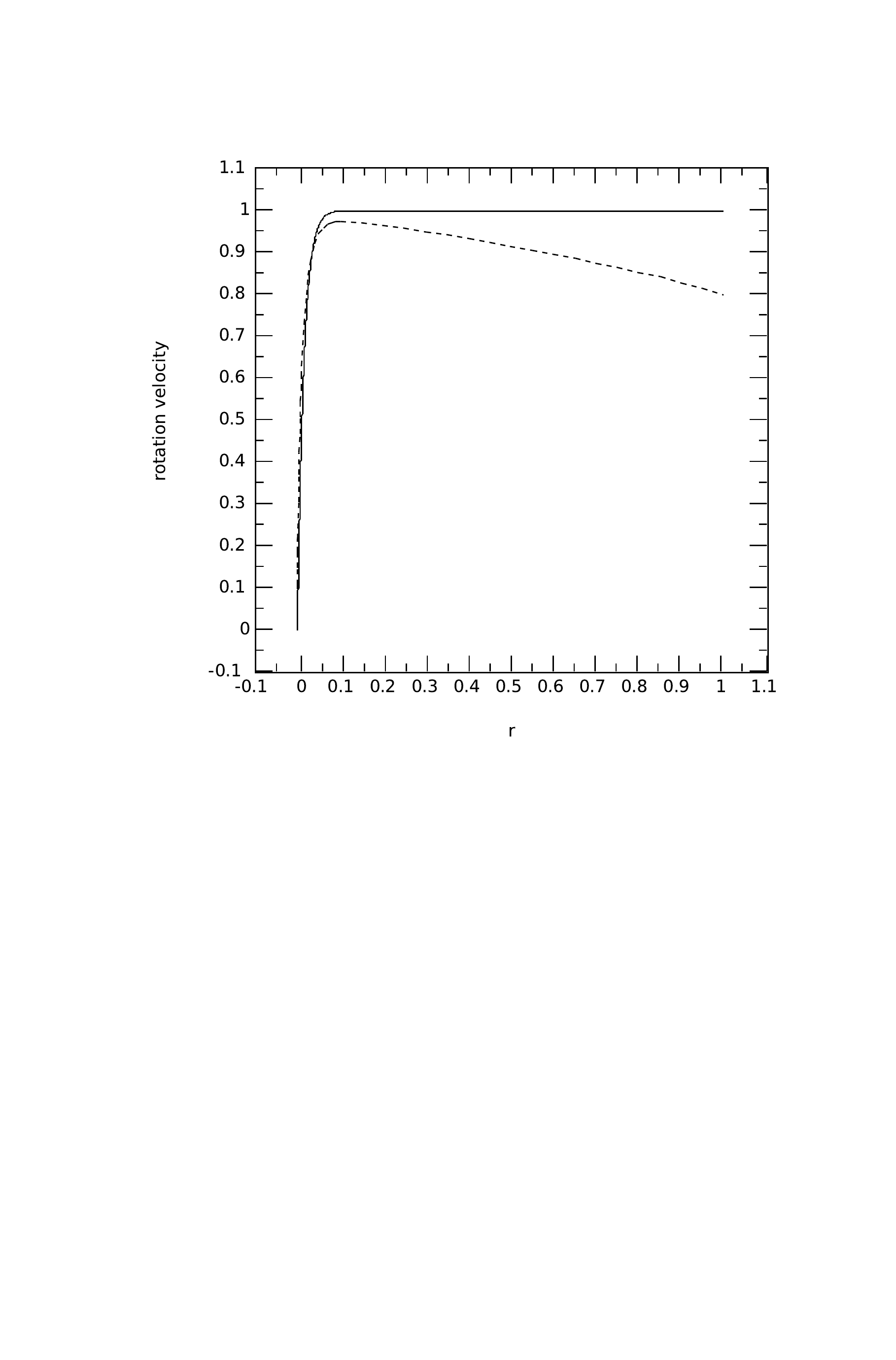}}
\caption{\label{fig:V-comparison}The rotation curve 
of $R_c = 0.015$ described by (\ref{eq:rotationCurves}) (solid) 
from Newtonian dynamics and that determined from (\ref{eq:force-balance?}) 
based on the Keplerian dynamics 
with the known $\rho(r)$ for $R_c = 0.015$ obtained in \S~3
with $A = 1.5714$  
(dotted).}
\end{figure*}

On the other hand,
if the mass density $\rho(r)$ is known as obtained in \S~3
for the case of $R_c = 0.015$ with $A = 1.5714$,
we can compute the corresponding $V(r)$ 
from (\ref{eq:force-balance?}).
figure~\ref{fig:V-comparison} shows that the $V(r)$ from
(\ref{eq:force-balance?}) clearly differs from 
that of (\ref{eq:rotationCurves}), with rotation velocity 
decreasing with the radial distance $r$ 
toward the galactic periphery as expected from 
Kepler's third law.
Again, this suggests that 
(\ref{eq:force-balance?}) according to the Keplerian dynamics 
cannot be a good approximation to the actual galactic dynamics.

As demonstrated in \S~3, we believe that Newton's laws 
of motion and gravitation can 
adequately describe the dynamical behavior of 
(disk) galaxies,
with appropriate mathematical treatments.
Kepler's laws should not be regarded as the same as Newton's laws.
Newton's laws can explain Kepler's laws for the planets in the solar system;
but Kepler's laws cannot be extended to the galactic dynamics
like Newton's laws, 
even in an approximation sense as shown in 
figure~\ref{fig:rho-comparison} and figure~\ref{fig:V-comparison}.

\section{Conclusions}
By strictly applying Newton's laws,
the present computational model can yield mass distributions
from the observed galactic rotation curves,
as apparently quite consistent with the observed 
near exponential brightness distributions.
Thus through the eyes of Newton,
galaxies are nothing more than graviationally bound assemblies of 
massive objects 
that are governed by his same laws for the planet's motion 
in the solar system.
Although Newton's laws and Kepler's laws seem to yield 
the same results when they are applied to the planets in the solar system,
they can lead to quite different results when describing 
the stellar dynamics in disk galaxies 
(cf. figure~\ref{fig:rho-comparison} and figure~\ref{fig:V-comparison}).
If not careful, simply extending Kepler's laws to 
the disk galaxies 
with subtensively distributed mass
can suggest misleading conclusions.

As demonstrated in \S~4, 
substituting the computed mass density distribution $\rho(r)$
based on Newtonian dynamics
into the Keplerian force-balance equation 
(\ref{eq:force-balance?}) 
would yield a rotation curve with orbital velocity decreasing toward 
galactic periphery (see figure~\ref{fig:V-comparison}).
This had led many authors to believe that the visible mass 
in a galaxy cannot explain the observed flat rotation curve
\cite{Rubin06, Rubin07, Freeman06}.
Therefore, some authors 
have speculated that some kind of invisible matter called dark matter 
must exist in the galaxy \cite{Bennett07, Rubin06, Rubin07, Freeman06}.
Other authors believed modification of Newton's laws
to be needed \cite{Milgrom83}.
The fundamental problem here is that astronomers tend to determine 
the `visible' mass in a galaxy from the measured brightness 
based on an over-simplified 
{\em mass-to-light ratio} \cite{Freeman06}.
When the mass distribution so estimated did not generate the 
observed {\em flat} rotation curve, especially when the Keplerian dynamics
was used for another over-simplification,
it is often referred to as the `galactic rotation problem' 
suggesting that there is a discrepancy between the observed rotation 
speed of matter in the disk and the predictions of Newtonian dynamics
\cite{Bennett07, Rubin06, Rubin07, Freeman06, Milgrom83}.
But we believe that Newtonian dynamics can adequately explain 
the stellar dynamics in disk galaxies when applied correctly;
neither the introduction of dark matter nor modification
of Newtonian dynamics is needed for 
explaining the observed rotation curve \cite{Gallo09, Gallo10}.
Hence, the rotating disk galaxies described with our model
must be that seen through the eyes of Newton.

\appendix
\section{Computational techniques}
Following a standard boundary element method \cite{Sladek98, Sutradhar08},
the governing equations (\ref{eq:force-balance}) and
(\ref{eq:mass-conservation})
can be discretized by dividing the one-dimensional
problem domain $[0, 1]$ into a finite number of line segments
called (linear) elements.
Each element covers a subdomain confined by two end nodes,
e.g., element $n$ corresponds to the subdomain
$[r_n, r_{n+1}]$, where $r_n$ and $r_{n+1}$ are nodal values of
$r$ at nodes $n$ and $n+1$, respectively.
On each element, which is mapped onto a unit line segment $[0, 1]$ in
the $\xi$-domain (i.e., the computational domain),
$\rho$ is expressed in terms of the linear basis functions as
\begin{equation} \label{eq:rho-xi}
\rho(\xi) = \rho_n (1 - \xi) + \rho_{n+1} \xi \, , \quad 0 \le \xi \le 1 \, ,
\end{equation}
where $\rho_n$ and $\rho_{n+1}$ are nodal values of $\rho$ at
nodes $n$ and $n + 1$, respectively.
Similarly, the radial coordinate $r$ 
(as well as $\hat{r}$) on each element is also expressed
in terms of the linear basis functions by
so-called isoparameteric mapping:
\begin{equation} \label{eq:r-xi}
r(\xi) = r_n (1 - \xi) + r_{n+1} \xi \, , \quad 0 \le \xi \le 1 \, .
\end{equation}

If $V(r)$ is given (e.g., from measurements),
the $N$ nodal values of $\rho_n = \rho(r_n)$ can be determined by
solving $N$ independent residual equations over $N - 1$ element
obtained from
the collocation procedure, i.e.,
\begin{equation} \label{eq:force-balance-residual}
\sum_{n = 1}^{N - 1} \int_0^1 \left[
\frac{E(m_i)}{\hat{r}(\xi) - r_i} - \frac{K(m_i)}{\hat{r}(\xi) + r_i}
\right]
\rho(\xi) h \hat{r}(\xi) \frac{d\hat{r}}{d\xi} d\xi 
+ \frac12 A V(r_i)^2
 = 0 \, , \, i = 1, 2, ..., N  \, ,
\end{equation}
with
\begin{equation} \label{eq:mi-def}
m_i(\xi) \equiv \frac{4 \hat{r}(\xi) r_i}{[\hat{r}(\xi) + r_i]^2} \, ,
\end{equation}
where $\rho(\xi) = \rho_n (1 - \xi) + \rho_{n+1} \xi$ and
$\hat{r}(\xi) = \hat{r}_n (1 - \xi) + \hat{r}_{n+1} \xi$.
The value of $A$ can also be solved by the addition of
the constraint equation
\begin{equation} \label{eq:mass-conservation-residual}
2 \pi \sum_{n = 1}^{N - 1} \int_0^1
\rho(\xi) h \hat{r}(\xi) \frac{d\hat{r}}{d\xi} d\xi - 1 = 0 \, .
\end{equation}
Thus, we have $N + 1$ independent equations for determining
$N + 1$ unknowns.
The mathematical problem is now well-posed.
The set of linear equations (\ref{eq:force-balance-residual}) and 
(\ref{eq:mass-conservation-residual})
for N + 1 unknowns, i.e., N nodal values of $\rho_n$ and $A$,
can be written in a matrix form as  
\begin{equation} \label{eq:matrix-form} 
\mbox{\boldmath $J$}\,\mbox{\boldmath $\cdot$}\, \mbox{\boldmath $x$} 
= - \mbox{\boldmath $R$} \, ,
\end{equation}
where $\mbox{\boldmath $R$}$ is the residual vector consisting of 
N + 1 components given by the 
left side of (\ref{eq:force-balance-residual}) and 
(\ref{eq:mass-conservation-residual}),
$\mbox{\boldmath $x$}$ is the unknown vector of 
N nodal values of $\rho_n$ and $A$,
and $\mbox{\boldmath $J$}$ is the Jacobian matrix of 
sensitivities of the residual $\mbox{\boldmath $R$}$ to the unknowns
$\mbox{\boldmath $x$}$, i.e., $J_{ij} \equiv \partial R_i/\partial x_j$. 
The matrix equation (\ref{eq:matrix-form}) is actually derived 
based on Newton's method (also known as the Newton-Raphson method)
for iteratively finding roots of a set of multi-variable nonlinear functions.
For a set of linear functions, as in the present case, 
a single iteration is enough for obtaining the solution.

The complete elliptic integrals of the first kind and second kind can
be numerically computed with the formulas \cite{Abramowitz72}
\begin{equation} \label{eq:K-m1}
K(m) = \sum_{l = 0}^4 a_l m_1^l - \log(m_1) \sum_{l = 0}^4 b_l m_1^l
\end{equation}
and
\begin{equation} \label{eq:E-m1}
E(m) = 1 + \sum_{l = 1}^4 c_l m_1^l - \log(m_1) \sum_{l = 1}^4 d_l m_1^l \, ,
\end{equation}
where
\begin{equation} \label{eq:m1-def}
m_1 \equiv 1 - m = \left(\frac{\hat{r} - r}{\hat{r} + r}\right)^2 \, .
\end{equation}
Thus, the terms associated with $K(m_i)$ and $E(m_i)$ in
(\ref{eq:force-balance-residual}) become singular when $\hat{r} \to r_i$
on the elements with $r_i$ as one of their end points.

The logarithmic singularity is treated by converting the
singular one-dimensional integrals into non-singular
two-dimensional integrals
by virtue of the identities:
\begin{eqnarray} \label{eq:log-integral-identities}
\left\{
\begin{array}{cc}
\int_0^1 f(\xi) \log \xi d\xi = - \int_0^1 \int_0^1 f(\xi \eta) d\eta d\xi
\\
\int_0^1 f(\xi) \log(1 - \xi) d\xi =
- \int_0^1 \int_0^1 f(1 - \xi \eta) d\eta d\xi  \,
\end{array}  \right . \, ,
\end{eqnarray}
where $f(\xi)$ denotes a well-behaving (non-singular) function of $\xi$
on $0 \le \xi \le 1$,
which can be derived by considering integration over a triangular area in
a two-dimensional $x\xi$-space, namely, 
\begin{eqnarray}
\int_0^1 f(\xi) \log \xi \, d\xi = 
\int_0^1 f(\xi)\left(\int_1^{\xi} \frac{dx}{x}\right)d\xi =
\int_0^1\left(\int_1^{\xi}\frac{f(\xi)}{x}dx\right)d\xi \nonumber \\
=-\int_0^1\left(\int_0^x\frac{f(\xi)}{x}d\xi\right)dx 
=-\int_0^1\frac{1}{x}\left(\int_0^1f(\eta x)\, xd\eta\right)dx
= -\int_0^1\int_0^1f(\eta \xi)\,d\eta d\xi \, , \nonumber
\end{eqnarray}
and 
\begin{eqnarray}
\int_0^1 f(\xi) \log(1- \xi) \, d\xi = 
-\int_0^1 f(\xi)\left(\int_0^{\xi} \frac{dx}{1-x}\right)d\xi =
-\int_0^1\left(\int_1^{1-\xi}\frac{f(\xi)}{y}dy\right)d\xi \nonumber \\
=-\int_0^1\left(\int_1^{1-y}\frac{f(\xi)}{y}d\xi\right)dy 
=-\int_0^1\frac{1}{y}\left(\int_0^1f(1-\eta y)\, yd\eta\right)dy \nonumber \\
= -\int_0^1\int_0^1f(1 - \eta \xi)\,d\eta d\xi \, . \nonumber
\end{eqnarray}

But a more serious non-integrable
singularity $1 / (\hat{r} - r_i)$ exists due to
the term $E(m_i) / (\hat{r} - r_i)$ in
(\ref{eq:force-balance-residual}) as $\hat{r} \to r_i$.
The $1 / (\hat{r} - r_i)$ type of singularity is treated by
taking the Cauchy principal value 
to obtain meaningful evaluation \cite{Kanwal96}.
In view of the fact that each $r_i$ is considered to be shared by two
adjacent elements covering the intervals $[r_{i-1}, r_i]$ and
$[r_i, r_{i+1}]$, the Cauchy principal value of
the integral over these two elements is given by
\begin{equation} \label{eq:CPV-def}
\lim_{\epsilon \to 0} \left[
\int_{r_{i-1}}^{r_i -\epsilon} \frac{\rho(\hat{r}) \hat{r} d\hat{r}}{
\hat{r} - r_i} 
+ \int_{r_i + \epsilon}^{r_{i+1}} \frac{\rho(\hat{r}) \hat{r} d\hat{r}}{
\hat{r} - r_i}\right]
\, .
\end{equation}
In terms of elemental $\xi$, (\ref{eq:CPV-def}) is equivalent to
\begin{eqnarray} \label{eq:CPV-xi}
-\lim_{\epsilon \to 0} \left\{
\int_0^{1 -\epsilon/(r_i-r_{i-1})} \frac{[\rho_{i-1} (1 - \xi) + \rho_i \xi]
[r_{i-1} (1 - \xi) + r_i \xi] d\xi}{
1 - \xi} \right . \nonumber\\
\left .
-\int_{\epsilon/(r_{i+1}-r_i)}^1 \frac{[\rho_{i} (1 - \xi) + \rho_{i+1} \xi]
[r_i (1 - \xi) + r_{i+1} \xi] d\xi}{
\xi} \right\}
\, .
\end{eqnarray}
Performing integration by parts on (\ref{eq:CPV-xi}) yields
\begin{eqnarray}
\rho_i\,r_i\,\log\left(\frac{r_{i+1}-r_i}{r_i-r_{i-1}}\right)-\left(
\int_0^1 \frac{d\{[\rho_{i-1} (1 - \xi) + \rho_i \xi]
[r_{i-1} (1 - \xi) + r_i \xi]\}}{d\xi} \log(1 - \xi) d\xi
\right . \nonumber \\
\left . +\int_0^1 \frac{d\{[\rho_{i} (1 - \xi) + \rho_{i+1} \xi]
[r_i (1 - \xi) + r_{i+1} \xi]\}}{d\xi} \log \xi d\xi
\right)
\, , \nonumber
\end{eqnarray}
where all the terms associated with $\log \epsilon$ 
cancel out each other, the terms with $\epsilon \log \epsilon$ 
become zero at the limit of
$\epsilon \to 0$.
The first term becomes nonzero when the mesh notdes are not 
uniformly distributed (namely, the adjacent elements are not of the 
same segment size).

At the galaxy center $r_i = 0$,
\[
\int_{r_i}^{r_{i + 1}} \frac{\rho(\hat{r}) \hat{r} d\hat{r}}{
\hat{r} - r_i} = \int_0^{r_{i + 1}} \rho(\hat{r}) d\hat{r} \, .
\]
Thus, the $1/(\hat{r} - r_i)$ type of singularity disappears naturally.

When $r_i = 1$ (i.e., $i = N$), it is the end node of the domain.
We can use a numerically relaxing boundary condition by
imagining another element extending beyond
the domain boundary covering an interval
$[r_i, r_{i+1}]$, because it is needed for
the treatment with 
Cauchy principal value.  
In doing so we can also have $r_{i+1}-r_i=r_i-r_{i-1}$ such that
$\log[(r_{i+1}-r_i)/(r_i-r_{i-1})]$ becomes zero, to
simplify the numerical implementation.
Moreover, it is reasonable to assume that $\rho_{i+1}=0$
because it is located outside the disk edge where
the extremely low intergalatic mass density is expected to have 
inconsequential gravitational effect.
With sufficiently fine local discretization, this extra element
can be considered to cover a diminshing physical space such that
its existence becomes numerically inconsequential.
Thus, at $r_i = 1$ we have
\begin{eqnarray}
\int_0^1 \frac{d\{[\rho_{i} (1 - \xi) + \rho_{i+1} \xi]
[r_i (1 - \xi) + r_{i+1} \xi]\}}{d\xi} \log \xi d\xi \nonumber \\
=  (\rho_{i+1}-\rho_i) \int_0^1 r(\xi) \log \xi d\xi
+ (r_{i+1}-r_i) \int_0^1 \rho(\xi) \log \xi d\xi 
= \rho_i\left[r_i - \frac32 (r_i - r_{i-1})\right] \, .
\nonumber
\end{eqnarray}
Now that only logarithmic singularities
are left, (\ref{eq:log-integral-identities}) can be used to eliminate
all singularities in integral computations.

In addition, to avoid cusps in mass density at the galactic center,
continuity of the derivative of $\rho$ at the galaxy center $r = 0$
is applied when solving for $\rho$ with given $V(r)$.
This boundary condition is imposed at the first node $i = 1$
to require $d\rho / dr = 0$ at $r = 0$, 
which becomes
\begin{equation} 
\rho(r_1) = \rho(r_2)
\, \nonumber 
\end{equation}
in discretized form.

Noteworthy here is that the (removable) singularities in the kernels of 
the integral equation (\ref{eq:force-balance}),
when properly handled, 
lead to a diagonally dominant Jacobian matrix in (\ref{eq:matrix-form})
with bounded condition number.
This fact makes the matrix equation (\ref{eq:matrix-form}) 
quite robust for almost any straightforward matrix solvers.  
In the present work, we simply used the available code 
for Gauss elimination \cite{Press88}.  
To check the correctness of our computational code implementation,
we substituted an exponential mass density distribution 
(e.g., $\rho(\hat{r})\, h = e^{-5\hat{r}}$)
into (\ref{eq:force-balance}) and compared the computed orbital velocity
$V(r)$ with the well-known analytical formula of Freeman \cite{Freeman70}.
The result showed excellent agreement.
Moreover, we could also obtain constant orbital velocity $V(r) = 1$ by 
integrating the density of Mestel's disk \cite{Mestel63}
$\rho(r) = A[1 - (2/\pi) \sin^{-1}(r)]/(2 \pi h r)$ 
through (\ref{eq:force-balance}).

The numerical method presented here can be applicable to 
rotation curves of arbitrary forms and   
does not require assumptions about the rotation curve beyond 
the radial coordinate where the orbital velocity is no longer measurable
as in  
Ref. \cite{Roberts75}
when using the formula of Toomre \cite{Toomre63} 
that contains 
an integral extending to infinity.
Not only is it convenient for considering galaxies of finite disk sizes,
it can also become an effective tool 
for deducing the mass distribution in a thin-disk galaxy from 
the measured rotation curve based on Newtonian dynamics. 

\ack
We are indebted to
Dr. Len Gray of Oak Ridge National Laboratory
for teaching detailed boundary element techiques for 
elegant removal of 
various singularities in integral equations.
We want also to thank Dr. Louis Marmet for his intuitive discussion
and preliminary computational results that convinced us to pursue
serious analysis of the galactic rotation problem.
The results shared by Ken Nicholson, Prof. Michel Mizony via 
similar approaches should be acknowledged for comfirming our beliefs.
Dr. Keith Watson's comments are appreciated for 
helping improve the presentation clarity.

\end{document}